 \documentclass[final,5p,times,twocolumn]{elsarticle}

\usepackage{lineno,hyperref}

\usepackage{xcolor}

\usepackage{epsfig}

\usepackage{amssymb}

\usepackage{float}

\journal{}

\begin{document}

\begin{frontmatter}



\title{Simulations for a low-perveance high-quality beam matching\\ of a high efficiency Ka-band klystron}

\author{M. Behtouei$^{a}$, B. Spataro$^{a}$, F. Di Paolo$^{b}$ and A. Leggieri$^{b}$\\\vspace{6pt}}

\address{$^{a}${INFN, Laboratori Nazionali di Frascati, P.O. Box 13, I-00044 Frascati, Italy};\\
 $^{b}${Dipartimento di Ingegneria Elettronica, Universit\`a degli Studi di Roma \textquotedblleft{Tor Vergata}\textquotedblright, Via del Politecnico, 1-00133-Roma, Italia }}

\begin{abstract}
Self consistent analytic and numeric design for a set of electron guns with a high beams quality to be used in  high power Ka-band klystrons are presented in this paper. The set of electron guns can be used in the high power Ka-band klystrons in order to feed linear accelerating structures at 36 GHz with an  estimated 20 MW  input power  by achieving an effective accelerating electric field in the (100-150) MV/m range. In the framework of the Compact Light XLS project, a short Ka-band linearizer by working at 36 GHz able for  providing an integrated voltage of at least 15 MV is proposed for bunch- phase linearization. In order to optimize the Ka-band klystrons efficiency for achieving 20 MW RF output power, different electron guns and beam focusing channel designs are examined and discussed in this paper.
\end{abstract}

\begin{keyword}
High Power Klystron,  Electron Gun, Particle Acceleration, Linear Accelerators, Free Electron Laser, Accelerator applications, Accelerator Subsystems and Technologies\end{keyword}

\end{frontmatter}



\section{Introduction}

The fundamental performance of all vacuum electron devices relies in the electron gun that, being the active device which produces the main electron stream to be manipulated, must respond to severe requirements and determines critically the feature of the power tube where it's applied to. In this study, the self consistent analytic design and the numeric model for a set of innovative electron guns suitable for Ka-band klystrons are described. For Ka band klystrons, very small beam dimension are needed and the presented family of electron guns responds to this requirement. Thermionic filament have been used to indirectly heat the presented cathode and a high vacuum of about $10^{-7}$ Torr is required for the correct operation.

The proposed electron guns are developed as beam sources for the Ka-band klystrons under development for the Compact Light XLS and also applyable to design a hard X-ray Free Electron Laser (FEL) facility using the latest concepts for bright electron photo injectors, very high-gradient X-band structures at 12 GHz, and innovative compact short-period undulators \cite{ref1}. The electron gun and the magnetic focusing system have to produce a 1 mm beam radius confined in a 1.2 mm pipe radius in order to obtain  about 20 MW beam power with a reliable operation. Estimations have been obtained by using  the numerical code  CST \cite{ref5} and analytical approaches. Also we are working on the design of the compact SW and TW accelerating structure operating on the third harmonic with a (100-125) MV/m accelerating gradient \cite{ref2,ref3,ref4} . 

Moreover, this activity is also of strong interest for the local activity in the framework of the Sparc-Lab project at INFN-LNF. 

Perveance, $K=I\ V^{-3/2}$ is the key element to control and measure the space charge force where I and V are beam current and voltage. This parameter doesn't allow to have identical velocity for each accelerated electrons due to the space charge effect. We are planning to finalize the linearizer structure design and the RF power source that will be able to produce up to 20 MW input power with an efficiency of about 42 $\%$ \cite{ref5,ref55,ref555}.  By considering that the efficiency is defined as the ratio of the output power to the input power, we conclude that the voltage is proportional to the ratio $(P_{out}/(\eta K))^{2/5}$. The high perveance leads to the strong space charge and consequently a weak bunching and the lower efficiency but if we intend to have a higher beam current we should increase the perveance. As a results, to maintain a good efficiency, we need to design an electron gun with a satisfactory perveance. 

In this paper, we present different electron guns and related magnetic devices in order to produce a 1 mm beam radius confined in a 1.2 mm beam pipe to maximize the klystron efficiency. The klystron works on the third harmonic of the bunched electron beam ($\sim$36 GHz) generated from a high-voltage DC gun (up to 480 kV). The cathode-anode geometry and the distance between them were optimized to adjust the electric field equipotential lines in order to obtain an extracted beam currents of above 100 A. 

In addition, the analytical approach for calculation of the electron gun's dimension has been accomplished and the results have been compared with numerical estimations obtained using CST Particle Studio.

\begin{figure*}[t]
\begin{center}
\begin{minipage}{38pc}
\includegraphics[width=18pc]{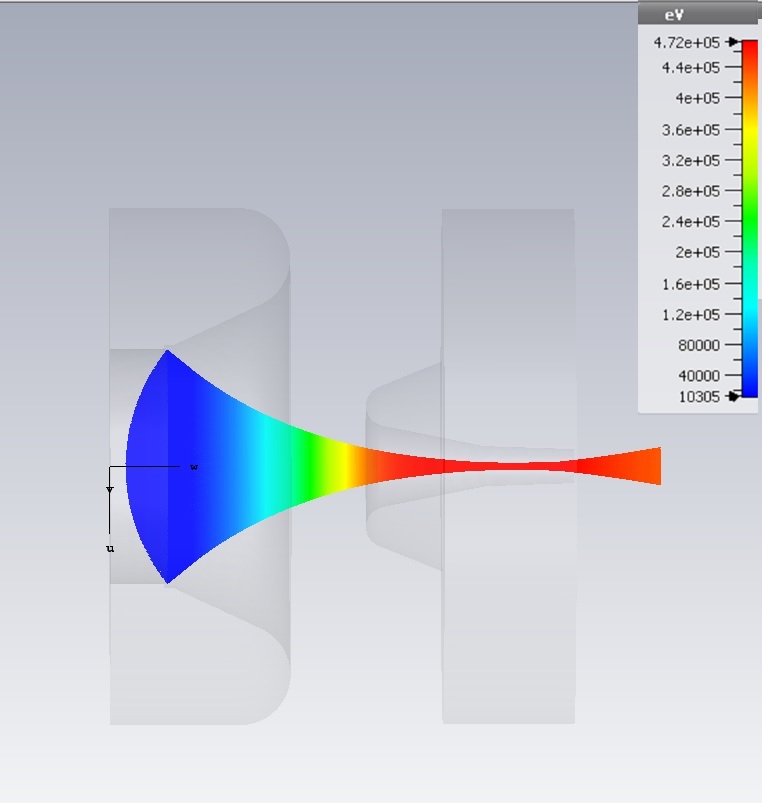}\hspace{1pc}
\includegraphics[width=21.3pc]{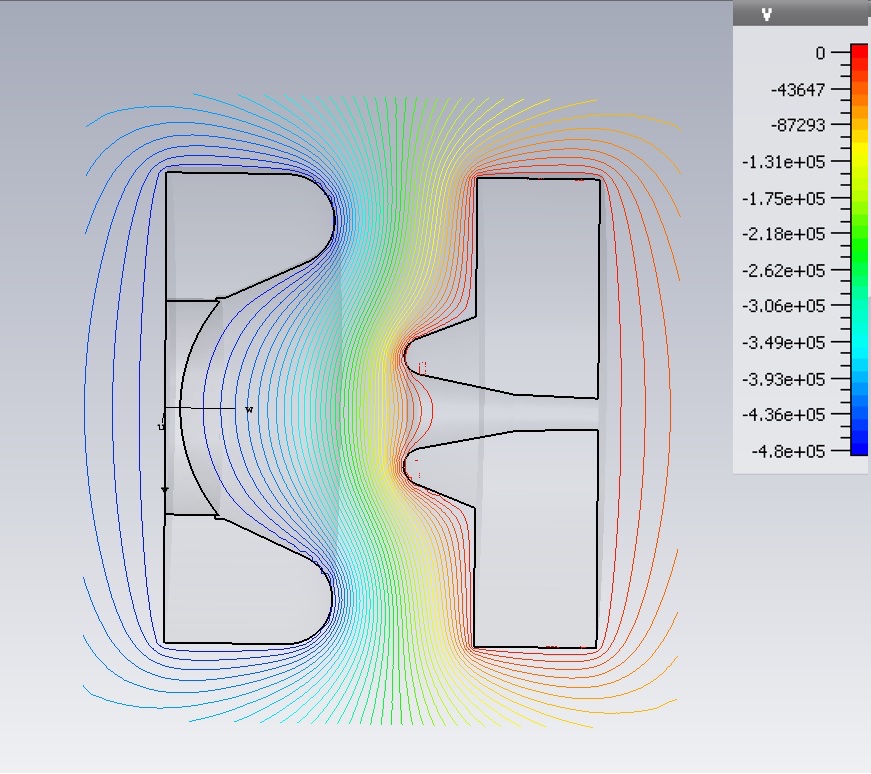}

\end{minipage} 
\end{center}
\caption{Preliminary electron gun design performed using the CST Particle Studio. The cathode-anode voltage is 480 kV and the beam current is 100 A. Beam trajectory (left) and Equipotential lines (right) are shown.}
\end{figure*}

\section{The Electron Gun injector design}

To feed the accelerating structure operating at Ka-Band (35 GHz), a Pierce-type electron gun as injector of the klystron have been designed. To produce a beam current of about 100 A with a beam power up to 48 MW, we need a cathode-anode voltage of about 480 kV.  In Fig. 1, the preliminary simulation of the electron gun with CST is shown. By manipulating the cathode-anode geometry, one can adjust the electric field equipotential lines in order to obtain a beam current extraction of 100 A. Beam trajectory (left) and electric field equipotential lines (right) are shown.  Design parameters of the diode gun for the Ka-band klystron are listed in Table 1.

\begin{table}[H]

\begin{center}
\caption{Design parameters of the diode gun for the Ka-band klystron}
\begin{tabular}{|| c| c||}
\hline
Design parameters&  \\ 
 \hline\hline
Beam power [MW]&  48\\  
\hline
Beam voltage [kV]& 480\\ 
\hline
Beam current [A]  & 100\\ 
\hline
 $\mu-$ perveance $[I/V^{3/2}]$&0.3 \\ 
 \hline
 Cathode diameter [mm]& 76 \\ 
\hline
Max EF on focusing electrode [kV/cm]& 200 \\ 
\hline
 Electrostatic compression ratio& 1488\\ 
\hline
\end{tabular}
\end{center}
\end{table}

The electrostatic beam compression ratio of 1488:1 has been obtained. The micro-perveance of the device is 0.3 $A V^{-3/2}$ and the maximum electric field on the focusing electrode is about 200 $kV/cm$ which is reasonable. In order to avoid the possible damage and safety operation margins in terms of pulse length, RF windows, power supply hardware stability, etc., we have decided to work with 480 kV at the cathode-anode voltage.

\section{Analytical method for estimating the dimensions of electron gun device}      
     
Writing the Poisson's equation in spherical coordinates and canceling out from the equation $\theta$ and $\phi$ because of symmetry about the axes, we may obtain the potential distribution between cathode and anode as follows,

\begin{equation}
\small{\frac{1}{r^2} \frac{\partial}{\partial r} (r^2 \frac{\partial V}{\partial r})+\frac{1}{r^2 \sin \theta}\frac{\partial}{\partial \theta} (\sin \theta \frac{\partial V}{\partial \theta})+\frac{1}{r^2 \sin^2\theta}\frac{\partial^2V}{\partial \varphi^2}=-\frac{\rho}{\epsilon_0}}
\end{equation}

\begin{figure*}[t]
\begin{center}
 \includegraphics[width=0.35 \textwidth ]{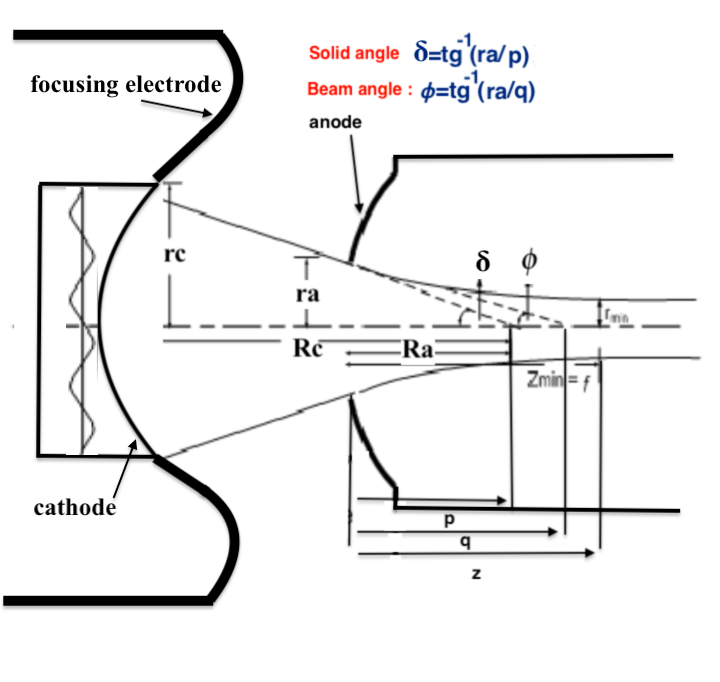}(a)
  \includegraphics[width=0.35 \textwidth ]{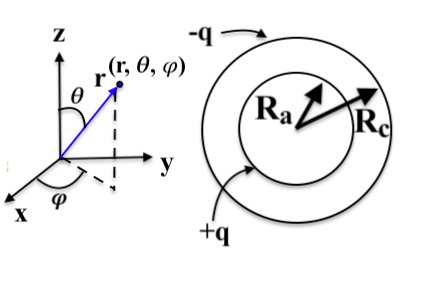}(b)
  
  \end{center}

\caption{a) Schematic view of the Pierce-type gun geometry  b) Two concentric conducting spheres of inner and outer radii $R_a$ and $R_c$, equivalent with the DC gun}
\end{figure*}

\begin{equation}
\small{\frac{1}{r^2} \frac{\partial}{\partial r} (r^2 \frac{\partial V}{\partial r})=-\frac{\rho}{\epsilon_0}=\frac{I}{4\pi r^2 \nu \epsilon_0}}
\end{equation}
 
where $\nu$ is the electron velocity. The final solution of the above equation is given \cite{ref6,ref7},

\begin{equation}\label{4}
\small{I=\frac{4 \epsilon_0}{9} (\frac{-2e}{m})^{1/2} \frac{V^{3/2}}{(-\alpha)^2}}
\end{equation}

where,

\begin{equation}\label{5}
\small{\alpha= \xi-0.3\ \xi^2+0.075\ \xi^3-0.0143\ \xi^4+....}
\end{equation}

and m is the electron mass, $\xi=log(\frac{R_a}{R_c})$  where $R_a$ and $R_c$ are the radii of the spheres of anode and cathode, respectively.  (see Fig. 2) \cite{ref8},

From Eq. ($\ref{4}$), the beam current is proportional to 3/2 power of the cathode voltage and the constant of proportionality is the perveance. The other parameters like the beam angle $\phi$ ($\phi=tg^{-1}(r_a/q)$)  can be obtained from electrostatic lens effect by considering analogy between light and charged-particle optics  \cite{ref9} (see Fig.(2)).

 \begin{table}[h]
\caption{Comparison among analytical and numerical results for estimating the dimensions of the electron gun device}
\begin{center}
\small{\begin{tabular}{||c|c|c||} 
\hline
Parameters& Analytical& Numerical (CST) \\ 
 \hline\hline
$\frac{R_c}{R_a}$&  3.18&3.125\\ 
\hline
$r_a$ [mm]&11.64&13.10 \\ 
\hline
Solid angle, $\delta$& 15.96$^\circ$&18.13$^\circ$\\ 
\hline
Beam angle, $\phi=tg^{-1} (r_a/q)$&13.11$^\circ$ &14.68$^\circ$\\ 
\hline
 Beam current [A]  & 100&100\\ [0.5ex] 
 \hline
\end{tabular}}
\end{center}
\end{table}

In Table 2, we compared the analytical and numerical results for estimating the dimensions of the electron gun device obtaining a good agreement.

\section{Magnetostatic Simulation}

After the electron gun exit, the beam propagates inside the beam pipe and the transverse dimension of the beam due to the existence of space charge  increases. This phenomena has more effect for the high current beam (of the order of hundreds of amperes) compared to the low current. To avoid the increase of the transverse dimension, a transverse focusing magnet have to be considered. To achieve the required compression of the beam after existing the electron gun, the magnetic field distribution has been analyzed. The 3D model of the gun and of the beam pipe is shown in Fig (3a). In the large beam pipe, we can arrange some slots to improve the vacuum pumping. On the other hand, the buncher and the output cavities should be installed on the small beam pipe. The magnetic field profile is shown in the Fig.  (3c) and the corresponding beam envelope in Fig.  (3b). We reported the design parameters of the gun with the focusing magnetic field along the beam axis in Table 3.

\begin{table}[h]

\caption{Design parameters of the gun with the focusing magnetic field along the beam axis}

 \begin{center}
\begin{tabular}{||c|c|c||} 
\hline
Design parameters&  \\ 
\hline\hline
Beam power [MW]&  48\\ 
\hline
Beam voltage [kV]& 480\\  
\hline
Beam current [A]  & 100\\ 
\hline
 Micro-perveance $[I/V^{3/2}]$&0.3\\  
 \hline
 Cathode diameter [mm]&  76 \\ 
 \hline
  Pulse duration [$\mu$ sec]& 0.1 \\ 
  \hline
Minimum beam radius in magnetic system [mm]& 0.98\\ 
\hline
Nominal radius [mm]& 1.00\\ 
\hline
 Max EF on focusing electrode [kV/cm]&  200 \\ 
 \hline
 Electrostatic compression ratio& 1488:1\\ 
 \hline
Beam compression ratio& 1635:1\\ 
\hline
 Emission cathode current density [$A/cm^2$]&2.02\\ 
 \hline
Transverse Emittance of the beam [mrad-cm]& 1.23 $\pi$\\ 
\hline
\end{tabular}
\end{center}
\end{table}

\begin{figure*}[t]

 \begin{center}
  \includegraphics[width= 0.95\textwidth ]{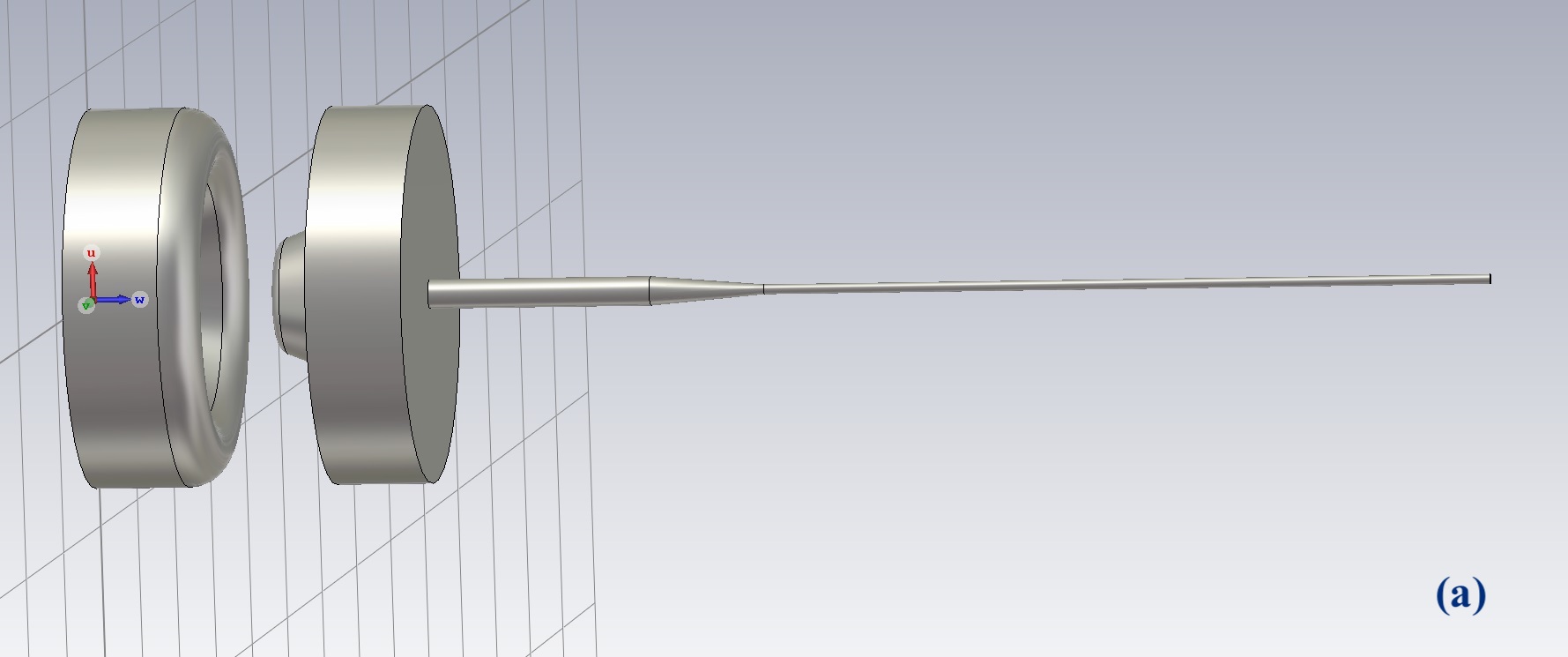}
    \includegraphics[width= 0.52\textwidth ]{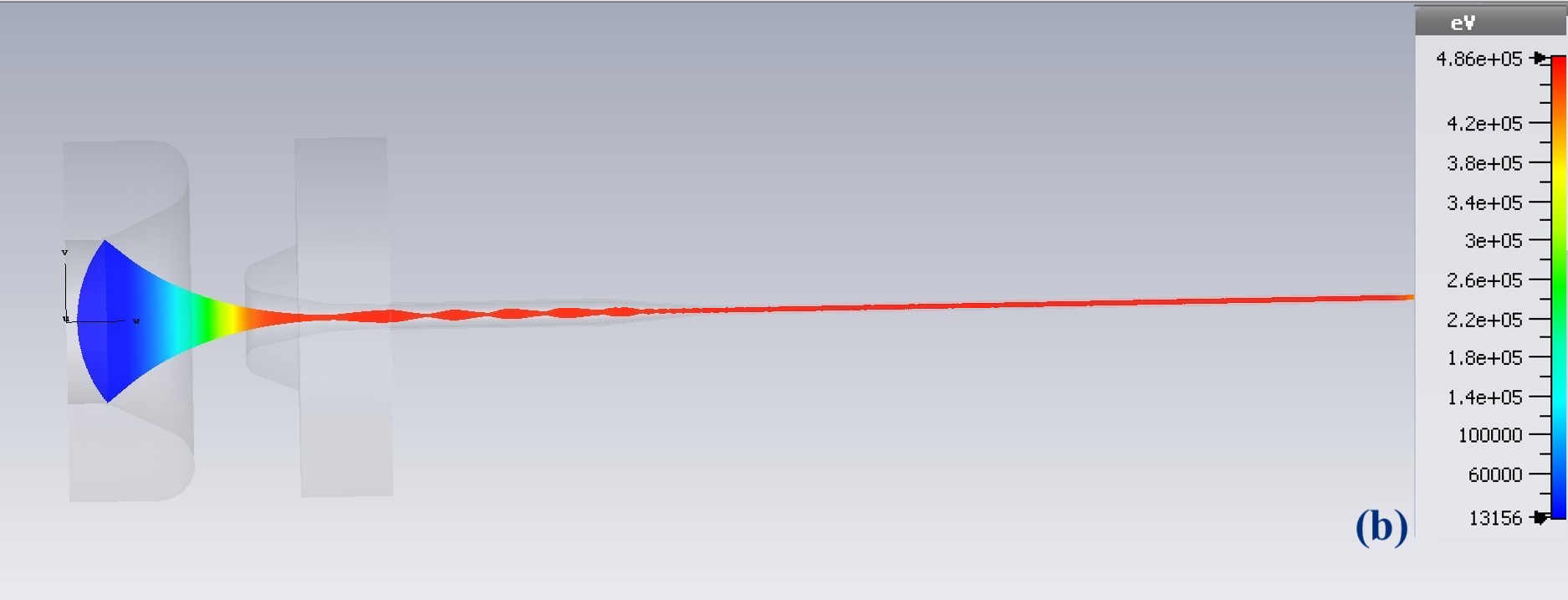}
      \includegraphics[width=0.42 \textwidth ]{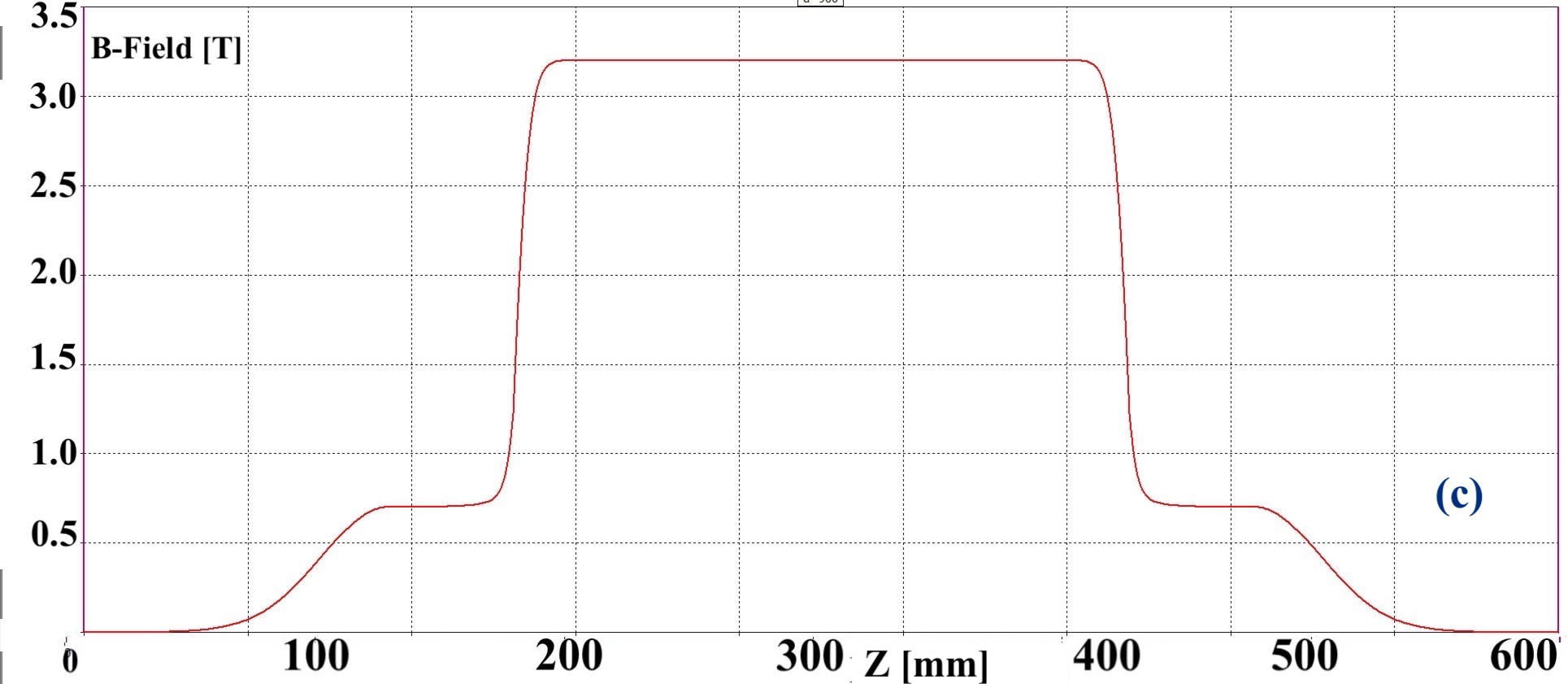}

\caption { a) 3D model of the gun and of the beam pipe. The cathode-anode voltage is 480 kV and the beam current 100 A b) Beam trajectory along the propagation direction c)  the axial magnetic field distribution. }

 \end{center}

      \end{figure*}

\begin{figure*}[t]

 \begin{center}
 \includegraphics[width=0.49 \textwidth ]{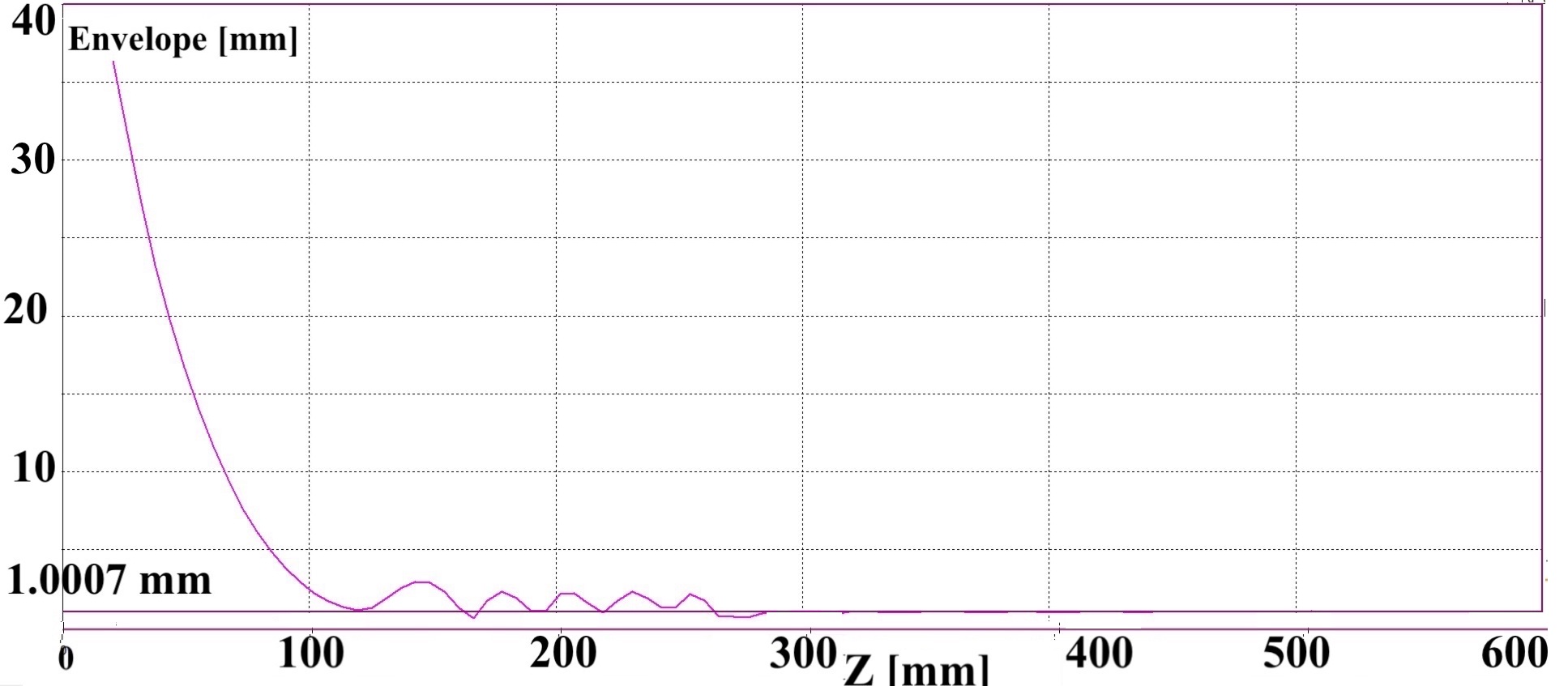}
    \includegraphics[width=0.49 \textwidth ]{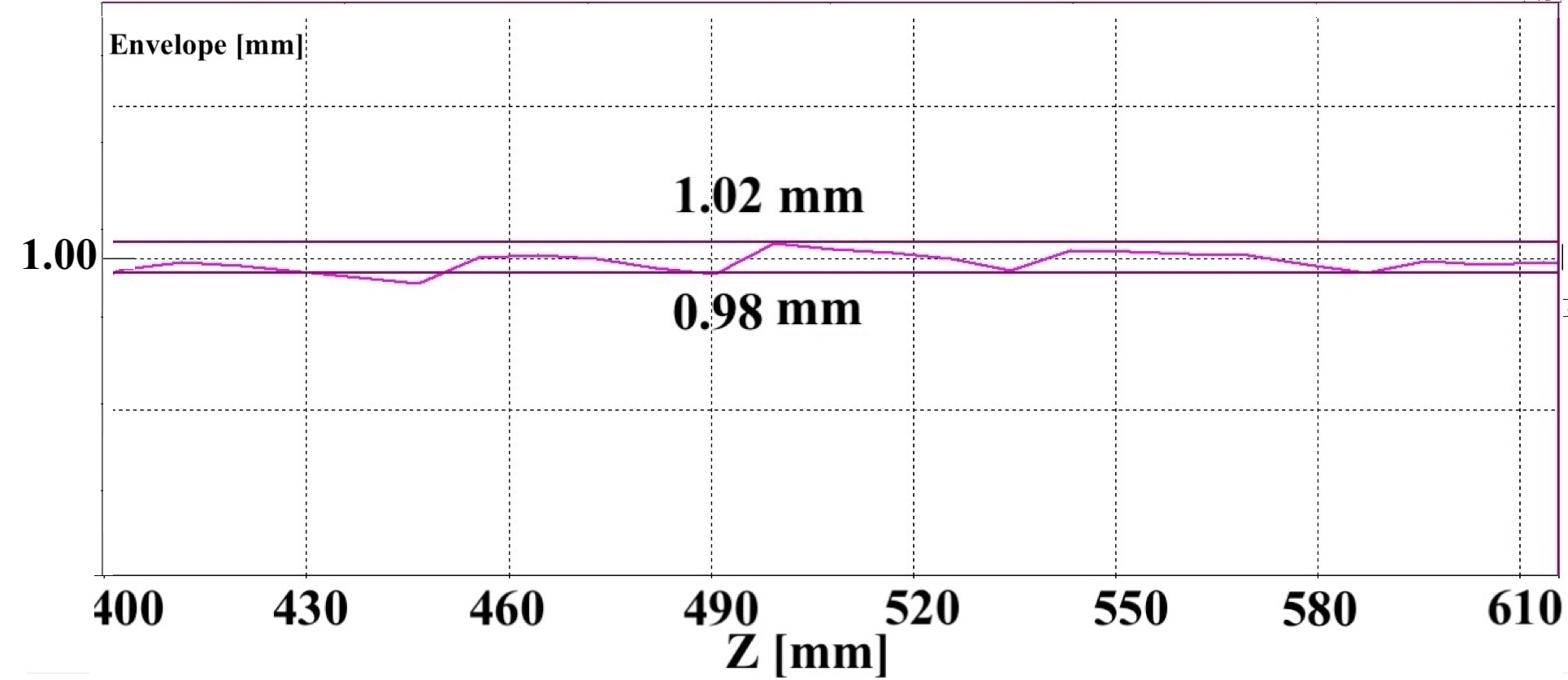}
       \includegraphics[width= 0.8\textwidth ]{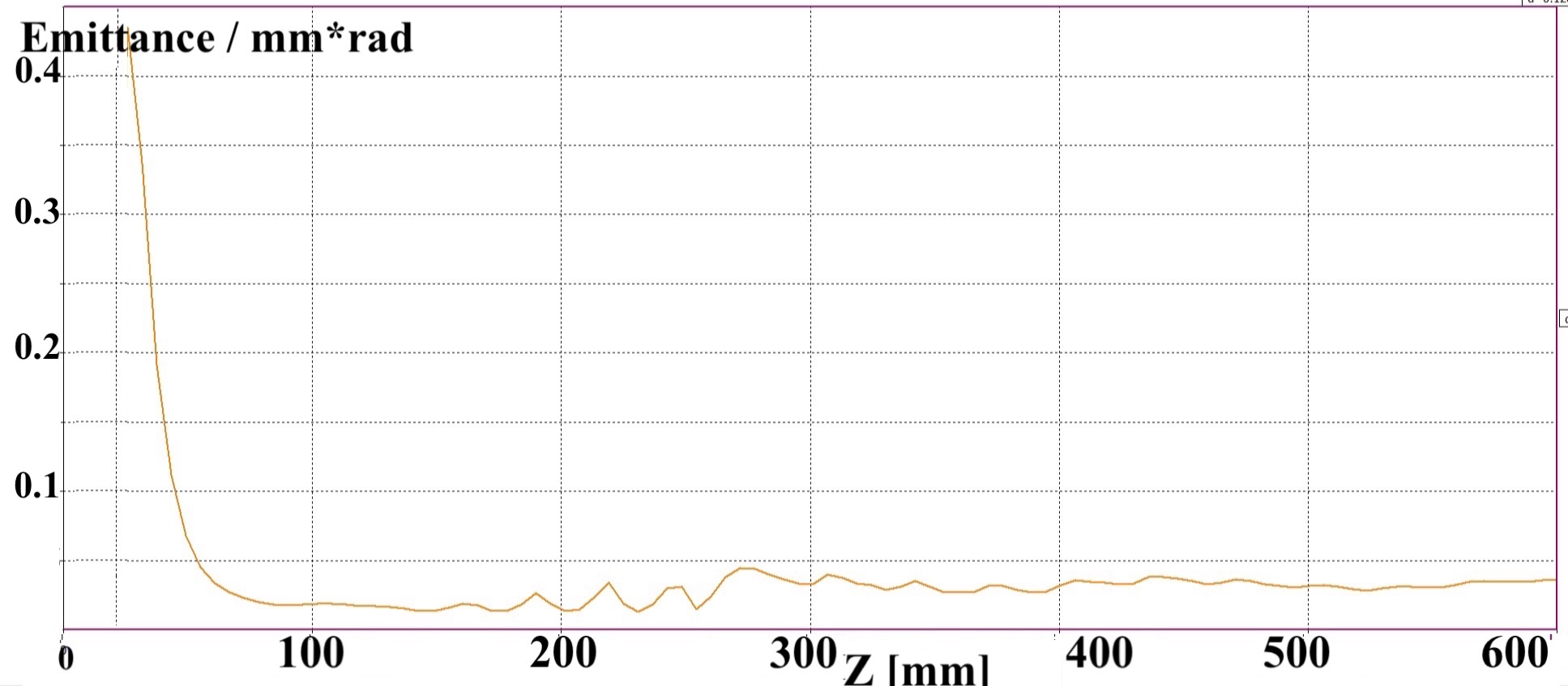}
   
 \end{center}

\caption{\footnotesize a) The beam envelope  and b) the transverse emittance of the beam along the beam axis with the focusing magnetic field. The transverse emittance rises in the small beam pipe where we have the minimum beam radius. c) the beam envelope inside the small pipe.}

     \end{figure*}

The magnetic field profile, shown in Fig. (3b), has a small peak of 7 kG and a constant magnetic field of 32 kG along a distance of  about 300 mm to obtain a narrow beam radius to allow cavities operate on the third harmonic of the fundamental mode of X-band. These cavities work in the Ka-band regime and therefore they require a small beam radius of about 1 mm. The design parameters of the gun with focusing magnetic field along the beam axis are listed in Table 4. In the region before the small pipe,  where we have mounted a tapered pipe, the magnetic field is 7 kG and consequently the beam radius is $\sim$ 2.31 mm but a value considerably higher than Brillouin limit of 0.6 mm. Likewise for the region where the field is 32 kG, the beam radius is $\sim$ 1 mm which again is much larger than the Brillouin limit of about 0.1312 mm. 
In order to increase the efficiency, we have to optimize the perveance. As we have mentioned before, efficiency and perveance are inversely proportional to each other \cite{ref10}. This means, in order to obtain a high efficiency, the perveance should be small. The price to pay is to have a small beam current. For this reason and also for other reason we will discuss in the next sections, we decided to work with a beam current of 100 A. The corresponding micro-perveance is 0.3 $A/V^{3/2}$. It should be noted that the micro-perveance of 0.657 $A/V^{3/2}$ is a common value for a modern klystron that can work up to about 235 A, but with an efficiency much smaller with respect to 100A operation. To change the micro-perveance from 0.657 $A/V^{3/2}$  to 0.3 $A/V^{3/2}$, we left the cathode-anode shapes unchanged but increased the distance between them. As shown in section 3, by increasing the cathode-anode distance, the ratio $\xi=log(R_a/R_c)$ decreases. Then, $\alpha$ (see Eq. (\ref{5})) will be higher and consequently the current obtained from Eq. (\ref{4}) becomes smaller. The reason for which we didn't change the cathode radius to decrease the current is because of intolerable increase of the maximum electric field on focusing electrode and a much smaller electrostatic compression. Instead, by increasing the cathode-anode distance, we obtain higher electrostatic compression ratio, 488, and a smaller electric field on the focusing electrode $\sim$ 20 MV/m. Hence, the electrostatic compression ratio and maximum electric field on the focusing electrode for the micro-perveance of 0.657 $A/V^{3/2}$ are 210 and 24 MV/m, respectively. 

The maximum possible beam compression is necessary to avoid the voltage breakdown \cite{ref11}. To increase the beam compression (minimizing the beam radius), the transverse emittance rises as we shown in Fig 4. We observe that the scalloping effect should be kept within 2$\%$ in order to optimize the klystron efficiency \cite{ref12}.
 
The magnetostatic beam compression ratio of 1635:1 is obtained where the beam radius $\sim$ 1 mm is located. Actually, the maximum possible compression ratio occurs when the beam radius reaches to the Brillouin limit. The transverse emittances of the beam in the small pipe is 1.23 $\pi$ (m\ rad-cm. 

For the new family of high power klystrons operating in the Ka-band under development, two other electron guns have been designed. One of them operates at the very high current (218A) and the other at the lower current of 50A.
The design parameters of the gun with the focusing magnetic field along the beam axis for the beam current of 50A are listed in Table. 4.
\begin{table}[h]

\caption{Design parameters of the gun with the focusing magnetic field along the beam axis}

 \begin{center}
\begin{tabular}{||c|c|c||} 
\hline
Design parameters&  \\ 
\hline\hline
Beam power [MW]&  24\\ 
\hline
Beam voltage [kV]& 480\\  
\hline
Beam current [A]  & 50\\ 
\hline
 Micro-perveance $[I/V^{3/2}]$&0.15\\  
 \hline
 Cathode diameter [mm]&  45.5 \\ 
 \hline
  Pulse duration [$\mu$ sec]& 0.1 \\ 
  \hline
Minimum beam radius in magnetic system [mm]& 0.82\\ 
\hline
Nominal radius [mm]& 0.925\\ 
\hline
 Max EF on focusing electrode [kV/cm]&  210 \\ 
 \hline
 Electrostatic compression ratio& 106:1\\ 
 \hline
Beam compression ratio& 697:1\\ 
\hline
 Emission cathode current density [$A/cm^2$]&2.67\\ 
 \hline
Transverse Emittance of the beam [mrad-cm]& 0.63 $\pi$\\ 
\hline
\end{tabular}
\end{center}
\end{table}

\begin{figure}[H]
\begin{center}

\includegraphics[width=0.35 \textwidth]{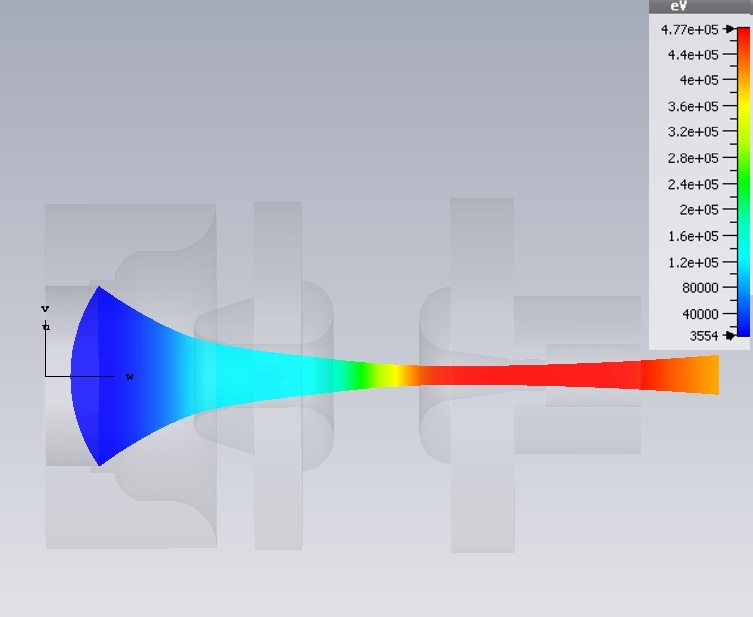}
\includegraphics[width=0.35 \textwidth]{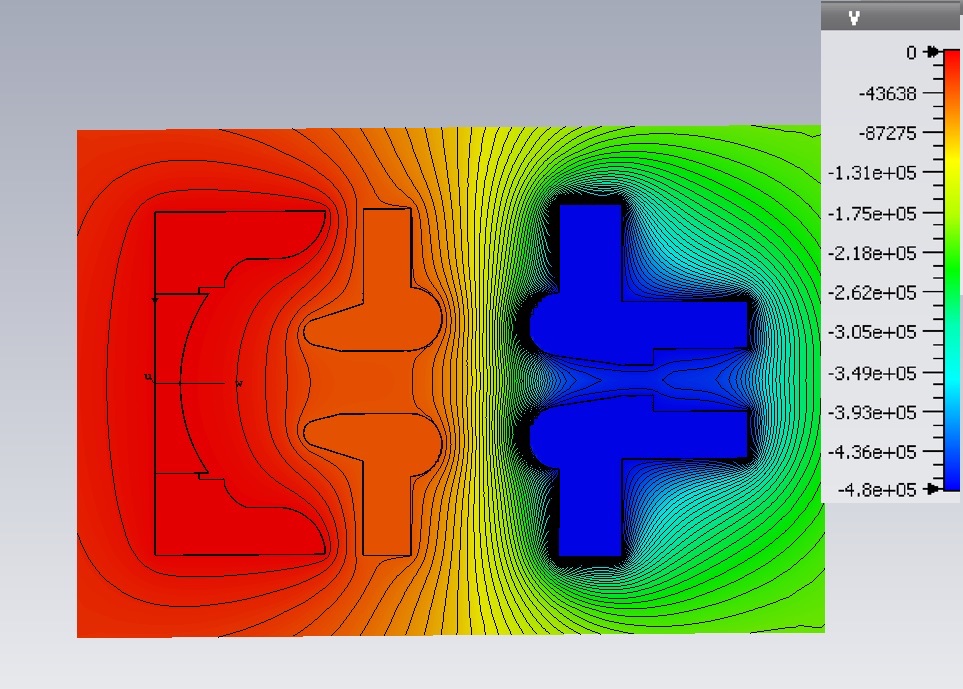}

\end{center}
\caption { Preliminary two anodes electron gun design from CST.  The cathode-anodes geometry was optimized to adjust the electric field equipotential lines in order to obtain a beam current extraction of 50A for the cathode-anode voltage of 480kV. Beam trajectory (left) and Equipotential lines (right) are shown. }

      \end{figure}

 The beam trajectory (left) and the electric field equipotential lines (right) are shown in Fig. (5). The cathode-anodes geometry was optimized to adjust the electric field equipotential lines to obtain a beam current extraction of 50A. Fig.s (6a), (6b) and (6c)  show the 3D model of the gun and of the beam pipe, the beam trajectory along the propagation direction and the axial magnetic field distribution, respectively. 
 
In order to investigate how we can reduce the axial magnetic field distribution along the beam propagation, we used a two anode configuration and the intermediate anode is being used for this purpose.  Like the previous system with a current electron gun operating at 100 A, the beam radius is set to 1 mm in a beam pipe of 1.2 mm. The magnetic field needed to compress the beam is reduced from 3.2 T to 0.7 T with this type of electron gun. 
\begin{figure*}[t]

 \begin{center}
 \includegraphics[width=0.8 \textwidth ]{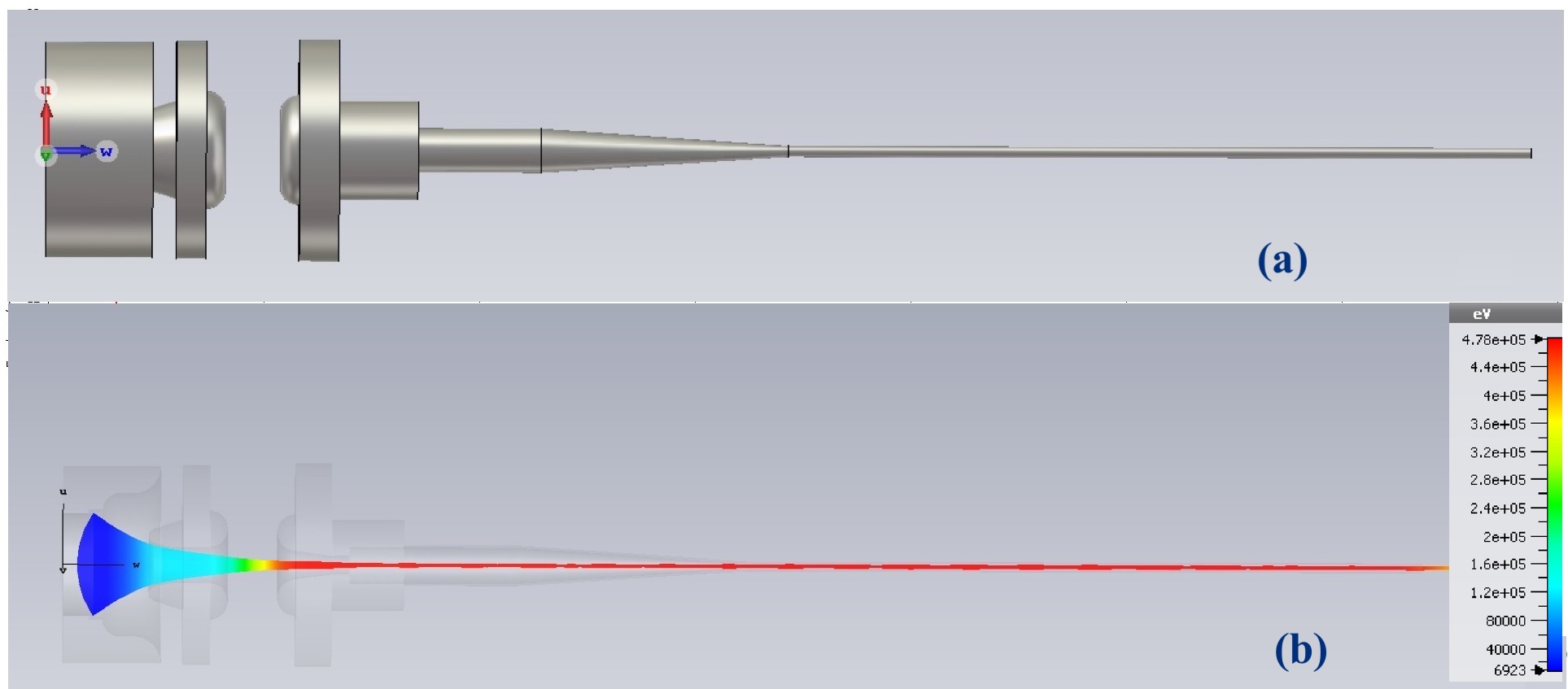}\\
   \includegraphics[width= 0.8 \textwidth ]{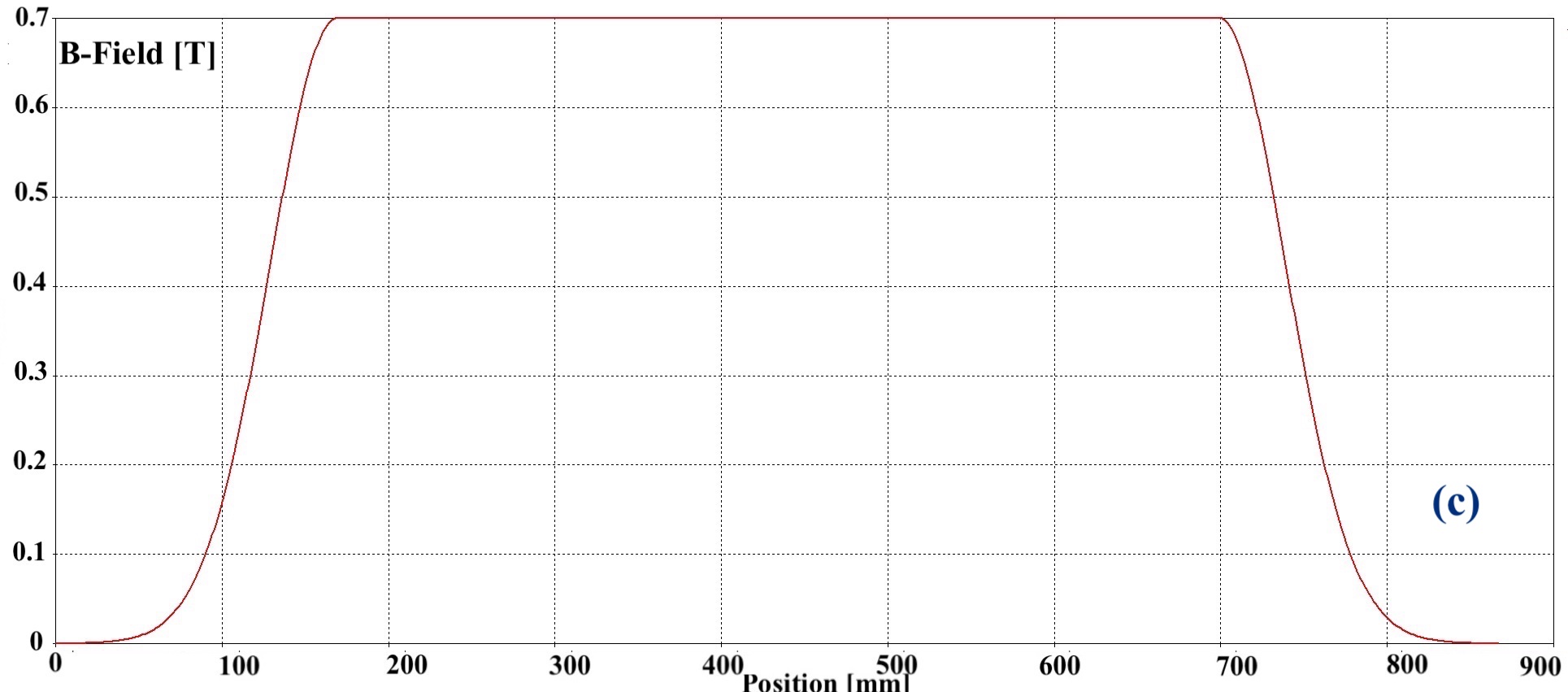}
 \end{center}

\caption { a) The 3D model of the gun and of the beam pipe. The cathode-anode voltage is 480 kV, producing a beam current of 50 A    b) the beam trajectory along the propagation direction c)  the axial magnetic field distribution. }

     \end{figure*}

The design parameters of the gun with the focusing magnetic field along the beam axis is reported in Table 5. The micro-perveance of this device is 0.15 $A V^{-3/2}$ which is half of the one with 100 A.  The maximum electric field on the focusing electrode is about 210 $kV/cm$ almost the same of the previous configuration with standard anode.

 In Table. 5 we list the design parameters of the gun with the focusing magnetic field along the beam axis for the current electron gun operating at 218 A . The design is the same as the gun operating at 100 A, but as mentioned in the previous section, the distance between cathode and anode is smaller and consequently the perveance is almost the double. 
 
 \begin{table}[H]

\caption{Design parameters of the gun with the focusing magnetic field along the beam axis}

 \begin{center}
\begin{tabular}{||c|c|c||} 
\hline
Design parameters&  \\ 
\hline\hline
Beam power [MW]&  105\\ 
\hline
Beam voltage [kV]& 480\\  
\hline
Beam current [A]  & 218\\ 
\hline
 Micro-perveance $[I/V^{3/2}]$&0.65\\  
 \hline
 Cathode diameter [mm]&  76 \\ 
 \hline
  Pulse duration [$\mu$ sec]& 0.1 \\ 
  \hline
Minimum beam radius in magnetic system [mm]& 0.92\\ 
\hline
Nominal radius [mm]& 0.99\\ 
\hline
 Max EF on focusing electrode [kV/cm]&  250 \\ 
 \hline
 Electrostatic compression ratio& 1488:1\\ 
 \hline
Beam compression ratio& 1635:1\\ 
\hline
 Emission cathode current density [$A/cm^2$]&2.67\\ 
 \hline
Transverse Emittance of the beam [mrad-cm]& 0.63 $\pi$\\ 
\hline
\end{tabular}
\end{center}
\end{table}

 \section{Conclusions}
In this paper, we estimated the dimensions of the electron gun device analytically and verified it with a numerical results performed using the CST code. Then, we accomplished electromagnetic and beam dynamics design of this innovative electron gun suitable for the Ka-band klystrons, by using the Microwave CST code. The electron flow is generated from a high-voltage DC gun (480 kV) and different beam currents ( 50A, 100A, and 218 A) have been extracted by changing the cathode-anode geometry in order to adjust the electric field equipotential lines. The electron beam is transported through the klystron channel and the beam confinement is obtained by means of a high magnetic field produced by superconducting coils. The channel has been optimized to deliver 24 MW, 48 MW and 105MW 
electron beams with a spot size of 2mm diameter for 50A, 100A, and 218 A respectively. In order to maximize the klystron efficiency we have decided to use the electron gun with the cathode voltage of 480kV and  the beam current of 50A and 100A.

In the forthcoming paper we will report the RF beam dynamics simulations.

\section*{Acknowledgment}
This work was partially supported by the Compact Light XLS Project, funded by the European Union's Horizon 2020 research and innovation program under grant agreement No. 777431.

\end{document}